# Thermodynamic Branch in the Chemical System Response to External Impact

B. Zilbergleyt[I].

INTRODUCTION.

Complex behavior of chemical systems is one of the more attractive and, at the same time, one of the most speculative topics of research for decades. The most recognized explanation of complex behavior takes its origin from kinetic models [1, 2]; thermodynamic approach to the problem arrived recently with discrete thermodynamics of chemical equilibria (DTd) [3]. This theory offers a more natural, simple, and clear picture of chemical system behavior under external impact, allowing us to follow a system response path from "true" thermodynamic equilibrium (TdE), back to its initial state, as well as in the opposite direction, along its forced advance beyond TdE to the logistic end of reaction, running within the system.

Let an isolated chemical system harbor one chemical reaction of synthesis from the group $aA+bB=A_aB_b$. Its basic state is TdE, where the internal thermodynamic force, thermodynamic affinity A, equals to zero. When the system becomes *clopen* (closed or open), its state is defined by a competition between thermodynamic forces - external, TdF (*F* in writing) and internal, the bound affinity B, which originates as a system reaction to external impact, and is defined similarly to A [3]

(1) $$B_j = -\Delta G_j/\Delta \xi_j,$$

where $\Delta\xi_j$ is the extent of the j-subsystem reaction, which has been forced out of TdE. The shift from the basic state, TdE, caused by TdF, is $\delta\xi_j = 1-\Delta\xi_j$ (further on in writing $\delta_j$ and $\Delta_j$) and

(2) $$B_j = -RT\{\ln[\Pi_j(\eta_j,0)/\Pi_j(\eta_j,\delta_j)]\}/(1-\delta_j).$$

Here $\Pi_j(\eta_j,\delta_j)$ is a traditional product of the mole fractions (aka reaction quotient), presented via $\delta_j$ and thermodynamic equivalent of transformation $\eta_j$ – ratio of any participant amount, chemically transformed in an isolated system on the way of its reaction *ab initio* to TdE, to its stoichiometric coefficient; $\eta_j$ is the system invariant at given stoichiometry, initial composition of the reacting mixture and reaction $\Delta G^0$, $\eta_j \leq 1$. The first term under logarithm is related to equilibrium constant as $\ln\Pi_j(\eta_j,0) = -K_j/RT$.

Now, due to its clopenness, we have a guided system, or merely a subsystem of the larger system. Presentation of the external force as a finite power series of the shift $\delta_j$ and assumption of a balance between the internal and external thermodynamic forces, which keeps up the reaction rate within the chemical system at zero, are leading us to the basic expression of DTd as logistic map [3]

(3) $$\ln[\Pi_j(\eta_j,0)/\Pi_j(\eta_j,\delta_j)] - \tau_j[\omega_0(\delta_j)(1-\delta_j) + \delta_j(1-\delta_j^\pi)] = 0,$$

where $\tau_j$ is the shift growth factor; $\omega_0(\delta_j)$ is the adjustment factor, allowing us to put all weights in the $\delta_j$ power series to unities; and $\pi$ is the system complexity factor, equal to the higher power in the series. All values in map (3) are dimensionless. Obviously, the first term in (3) is reduced by (-RT) and multiplied by $(1-\delta_j)$ bound affinity, the second represents the TdF; denominator in $\tau_j$ is also RT. As a whole, map (3) is the Gibbs' free energy change in clopen chemical systems, accounting for TdF.

Throughout this paper we will be focused on the relationship between the system shift from TdE and the external force, causing this shift. With such an approach, graphic solutions to map (3), obtained by Iterative simulation, are *dynamic* bifurcation diagrams; so far we are not able to get their images beyond bifurcation point due to double-stability within bifurcation area (Fig.1 and Fig.2). Their shape in general is very much similar to the *static* bifurcation diagrams, plotted in the system shift from TdE vs. shift growth factor coordinates in Fig.1 and Fig.2. Graphs in Fig.3 and in Fig.4, left, are the plottable parts of *dynamic* diagrams with the system shift as ordinate vs. dimensionless thermodynamic force (TdF), a ratio TdF/RT, |TdF|=kJ/mol, as abscissa. Curves in Fig.1 through Fig.4, left, are loci of the equilibrium points between the TdF$_j$ and B$_j$.

[I] System Dynamics Research Foundation, Chicago, USA, e-mail sdrf@ameritech.net.

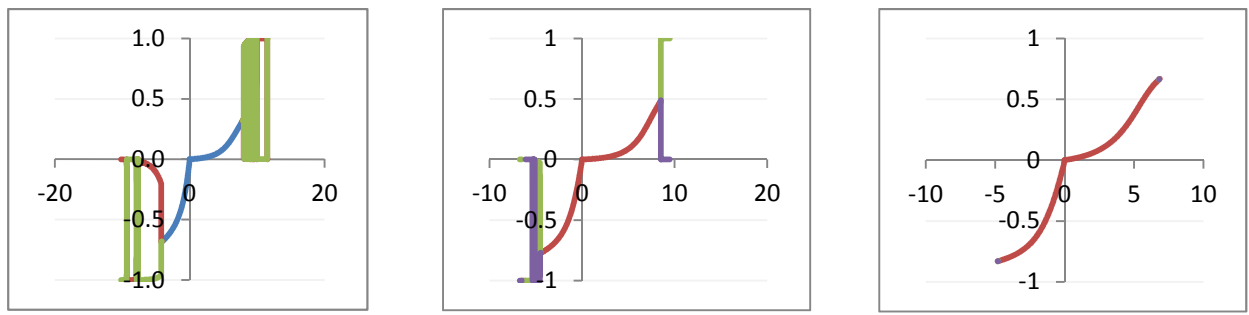

Fig.1. Graphical static solutions to map (1), δ (ordinate) vs. τ (abscissa), reaction A+B=AB; direct reaction in the I, reverse in the III quadrants, initial reacting mixture composition (1, 1, 0), moles, η=0.9933, (ΔG$^0$=-26.93, kJ/mol), varying complexity factor: π=1, 3, 6, left to right.

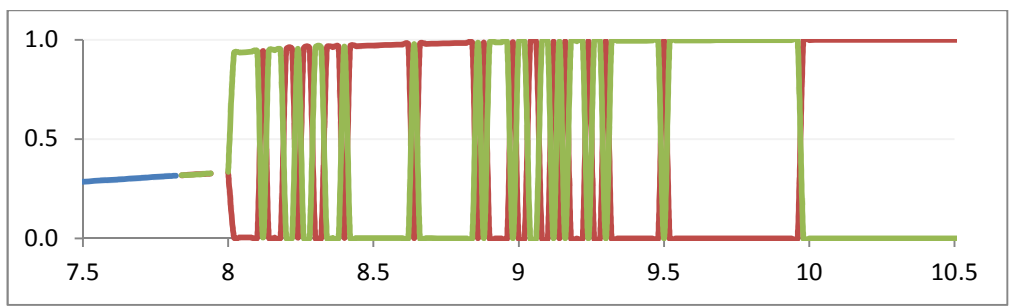

Fig.2. Rescaled image of the bifurcation area with oscillations, δ vs. τ, from the Fig.1, left, direct reaction.

As it is shown in Fig.1 and Fig.4, left, the curves start at TdE with δ$_j$=0 and $F_j$=0. As TdF increases, they pass the TdE area, adjacent to the zero point of the reference frame, the area of indifferent equilibria, where still δ$_j$=0, caused now by the system inertia, or the system resistance to external impact. Then follows the area of open equilibria, OpE, where δ$_j$ increases sharply from zero up to the bifurcation point at low π or, at high π, asymptotically approaching δ$_j$=1, which corresponds to the initial state of the system reaction. One can see in Fig.4, left, that extent of the TdE area along abscissa depends on η$_j$, contracting almost to zero at small η$_j$ values; η$_j$ is the measure of the system resistance to the external impact. Zone, formed by those two areas is known as thermodynamic branch (TdBra) [4], and so far this is the only part of dynamic diagrams that we can reproduce correctly. Practically, many - if not the most - of chemical systems do not experience bifurcations for many reasons, e.g., due to high value of the complexity factor, like the system with π=6 in Fig.1, right, and higher. However, being pushed by the external force away from TdE, all systems move along their TdBra, which is the first and often the only pattern of chemical systems response to external TdF. Its detailed investigation is the goal of this work.

SIMULATION RESULTS AND THEIR ANALYSIS.

A close look at the pictures in Fig.1 reveals the major difference between their TdBra – they end up at various shift values. Those are the partial TdBra. For the full TdBra image we had taken advantage of their changes with the system complexity factor, bringing $\pi$ up to extreme and perhaps non-realistic numbers of ≥150. Obtained this way curves are shown in Fig.3. As it follows from the left picture, all partial TdBra, obtained for the different $\pi$ values, are the parts of the full TdBra: no matter whatever is

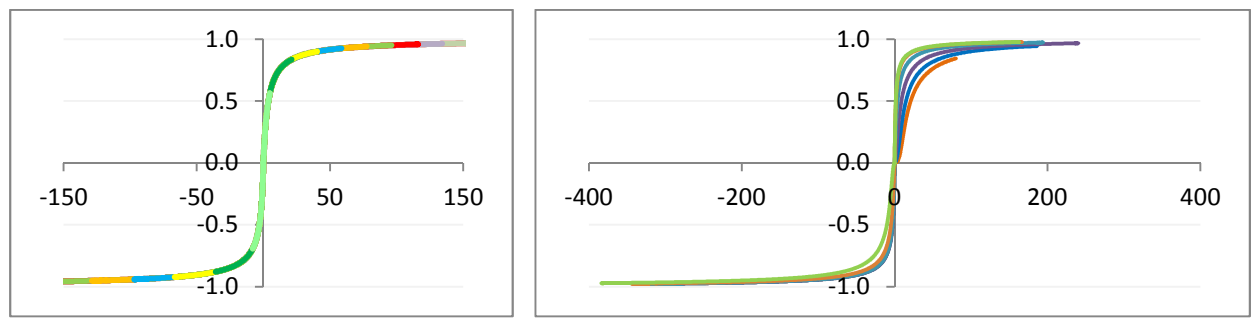

Fig.3. Full TdBra, $\delta$ (ordinate) vs. $F$ (abscissa). Left: different colors correspond to some partial curves, simulated at increasing $\pi$ values, $\eta_j$=0.6867. Right: set of full TdBra for different $\eta_j$, varying from 0.0094 (top curve) to 0.997 (lower curve), ($\Delta G_j^0$ from +10.66 to -31.18, kJ/mol). In both cases the reaction was A+B=AB, initial reacting mixture contained (1,1,0) moles of reactants.

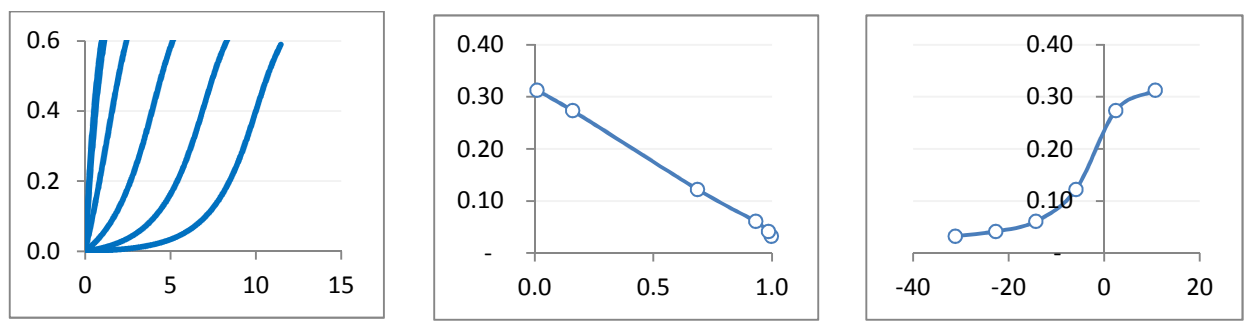

Fig.4. Left: rescaled initial area of the first quadrant from Fig.3, right, $\pi$=5, $\delta$ (ordinate) vs. $F$ (abscissa). Center: $d\delta/dF$ (ordinate) vs. $\eta$ (abscissa). Right: $d\delta/dF$ (ordinate) vs. $\Delta G^0$, kJ/mol, (abscissa). Reaction and conditions are the same as in Fig.3, right.

Table I.

| $\eta_j$ | $\Delta G_j^0$, kJ/mol | $d\delta_j/dF_j$ |
|---|---|---|
| 0.9970 | -31.18 | 0.0324 |
| 0.9854 | -22.70 | 0.0416 |
| 0.9305 | -14.28 | 0.0610 |
| 0.6857 | -5.92 | 0.1224 |
| 0.1594 | 2.40 | 0.2734 |
| 0.0094 | 10.66 | 0.3125 |

Table II.

| $\eta_j$ | $d\delta_j/dF_j$ |
|---|---|
| 0.997 | 0.03084 |
| 0.9854 | 0.03409 |
| 0.9305 | 0.04946 |
| 0.6857 | 0.118 |
| 0.1594 | 0.26537 |
| 0.0094 | 0.30737 |

the system complexity, it moves from TdE along the same curve, and the π value defines how far it will go. The right picture in Fig.3 shows a set of TdBra for various values of $\eta_j$: the bigger is $\eta_j$, the slower is the TdBra ascent, or the bigger is $\eta_j$, the stronger is the system resistance to TdF. The most interesting part of TdBra is the steepest one, it defines major differences between the system behavior at various $\eta_j$; a part of our task was to find empirical analytical dependence of the maximum slope steepness upon the system robustness. Rescaled parts of the full curves for direct reaction are shown in Fig.4, left. Apparently, the steepest parts of the curves within the $\delta_j$ range ~(0.1...0.5 and higher) may be approximated by the straight lines. We used this feature to find out the sought empirical analytical dependency. With the MS Excel tool LINEST, the values of $d\delta_j/dF_j$ have been estimated with regards to $\eta_j$ and $\Delta G_j^0$. Results occurred to be very close for different stoichiometric equations of the synthesis reactions; for $a$=1 and $b$=1 they are shown in Fig.4, center and right, appropriate numerical data are in Table I. As it follows from these pictures, dependency of the TdBra maximum steepness upon the thermodynamic equivalent of transformation $\eta_j$ is stronger than upon the reaction $\Delta G_j^0$. To obtain analytical relationship between $d\delta_j/dF_j$ and $\eta_j$, we have taken average (0.300±0.004) as the free term, and from the Table I and similar tables for various combinations of $a$ and $b$ in stoichiometric equations, with the LINEST, we have found that the average coefficient at $\eta_j$ equals to -(0.286±0.004), leading to following approximation

(4) $\qquad\qquad\qquad\qquad\qquad\qquad\qquad d\delta_j/dF_j = 0.30 - 0.29\eta_j.$

Calculated by this equation data for the A+B=AB reaction are in Table II, appropriate graphs for some reactions aA+bB=$A_aB_b$ with varying stoichiometric coefficients are shown in Fig.5, left. Interestingly, that even the chemical systems with non-stoichiometric initial reacting mixtures have shown the same, but slightly displaced linear dependency between $d\delta/dF$ and $\eta$, Fig.5, right. Presenting (4) in a rough outline

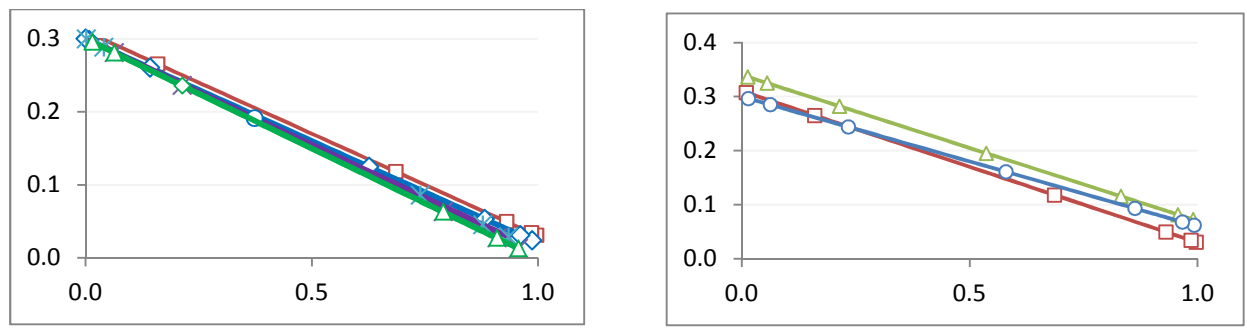

Fig.5. $d\delta_j/dF_j$ (ordinate) vs. $\eta_j$ (abscissa). Left: reactions A+B=AB (blue), A+2B=$AB_2$ (brown), 2A+3B=$A_2B_3$ (green), 5A+2B=$A_5B_2$ (purple), stoichiometric initial reacting mixtures. Right: reaction A+B=AB, A to B initial amounts ratio -  1:1 (brown), 1:2 (green), 1:3 (blue).

by neglecting the difference of only ≈3,4% between the free term and the coefficient at $\eta_j$, we can re-write and then integrate it, arriving at

(5) $\qquad\qquad\qquad\qquad\qquad\qquad\qquad \delta_j = 0.3(1-\eta_j)F_j.$

At $F_j$=0 we have also $\delta_j$=0, and integration constant is zero.

The term (1-$\eta_j$) draws some extra attention. If $\eta_j$ is a measure of the reaction completeness with regards to consumption of the reactants and formation of the products ($\eta_j$≤1 due to the law of simple proportions, and was always confirmed by simulation) and, in a sense, is also a measure of the reaction irreversibility, the term (1-$\eta_j$) may be considered a measure of the reaction reversibility – the more reversible is reaction, the less are the efforts needed to move it back to its initial state with $\delta_j$=1.

The meaning of these new empirical rules, based on computer simulation, is clear – the bigger is the reaction reversibility, the faster and the larger is the system escape from TdE at the same TdF. The pictures for the reverse reactions with $\Delta G_{jr}^0 = -\Delta G_{jd}^0$ and $\eta_{jr}$ are highly similar. Interestingly enough, logically it should be and always was confirmed by simulation, that $\eta_{jd}+\eta_{jr}$=1.

Because our results are relevant to the linear or approximated by straight line parts of the curves in Fig.3, an analogy to the well-known Hooke's law $F_j = -k\delta_j$ with $k=-[0.3(1-\eta_j)]^{-1}$ is clearly seen in (5). The number 0.3 is still unexplained.

## THERMODYNAMIC BRANCH AND CHEMICAL HYSTERESIS.

By definition, the symmetry is held unchanged along the TdBra area. That means also absence of hysteresis when the system returns to its initial state from any point on TdBra. Here is the logical proof of this statement.

Discrete thermodynamics considers equilibrium chemical (and quasi-chemical, e.g. see [5]) systems, either isolated or clopen, and is a general shell for equilibrium thermodynamics of chemical equilibria, including classical thermodynamics of chemical equilibria as a particular constituent. Indeed, if the increase rate of external TdF is comparable with the relaxation rate within the system to keep $B_j-F_j$=0, the system is moving from TdE along the *equilibrium* TdBra when TdF increases. For the isolated system, $F_j$=0 by virtue of its isolation, and only thermodynamic affinity, an internal thermodynamic force, controls the system, vanishing at TdE. Now, if in any point of the equilibrium TdBra the TdF growth stops and TdF starts to decrease or vanishes at all, the system turns from a guided into a free entity, and its move back to TdE is controlled exclusively by the bound affinity $B_j$, which now turns to regular affinity $A_j$. The system recedes along the same TdBra, which it followed when $B_j$ was increasing in order to balance the external force. Therefore, the system experiences no hysteresis (see the basic map derivation in [3]). As follows from the above as well as from appropriate simulation results, *the equilibrium path for the system of any type, isolated or clopen, from any point on TdBra back to TdE is the same as from TdE*.

## DISCUSSION.

TdBra was for a long time just a symbol of classicism in the theory of dynamic systems, whose image in the classical case of chemical systems might be guessed, but could not be drawn or investigated. It became possible only in the discrete thermodynamics of chemical equilibria [3]. Due to the unique opportunity of DTd to treat the TdBra quantitatively, the following features of its structure were discovered in this work:

1. The full, 2-way system TdBra looks like the S-shaped curve, with the flat ends asymptotically approaching limiting values of the shift: on the right side $\delta\to1$ and on the left side $\delta\to-1$, both correspond to initial states of the direct and the reverse reactions;
2. When the system complexity is varying and other system parameters are unchanged, all thermodynamic branches of the system follow the same path;

3. The dδ/d*F* dependence upon the system thermodynamic equivalent of transformation, η, is nearly linear within the restricted part of TdBra with maximum steepness, leading within this area to linearity of the system shift vs. TdF;
4. The reaction reversibility (1-η), which is an invariant given reaction stoichiometric equation, its ΔG$^0$ and initial reacting mixture composition, after multiplication by 0.3 equals to the maximum speed of the system deviation from TdE under impact of TdF; it serves as a proportionality coefficient between the system deviation from TdE and TdF that caused that deviation.
5. There is no difference, or hysteresis between the paths along the equilibrium TdBra to and from TdE in the free as well as in the guided chemical systems.

The empirical numbers in equations (4) and (5) have enough precision to exclude their occasional origin. There might be something strong, but so far not clear for us in the background of these data.

It occurred that the complexity factor π is reflecting correctly the real feature of chemical systems: the bigger is its value, the bigger must be the system shift from TdE to achieve the bifurcation area, and, beginning with a certain value of π, that area is not achievable (or does not exist) at all. This means that the greater is the system complicacy, the less it is prone to evolution, which sounds quite reasonable.

Map (3) is transcendent, not allowing us to express the shift as an analytical function of TdF, and we still are not aware of any experimental data that may be good for numerical analysis by the methods of discrete thermodynamics. Although this work may be considered abstract, its empirical observations, based on the computer experiment, offer some new practical possibilities in chemical engineering. For example, discovered linearity allows us to forecast behavior of real clopen chemical systems under external impact. Calculation of η is a simple routine computer task, and one can easily find a speed of chemical system deviation from TdE as well as compare the shifts at different η values in the same system or in different systems.